\documentclass[twocolumn,aps,pra,10pt,showpacs,floatfix]{revtex4-2}
\usepackage{bm}
\usepackage{amsfonts}
\usepackage{physics}
\usepackage{amssymb}
\usepackage{amsmath}
\usepackage{graphicx}
\usepackage{hyperref}
\hypersetup{
    colorlinks=true,
    linkcolor=blue,
    filecolor=blue,
    urlcolor=blue,
    citecolor=blue,
    }
\begin{document}
\title{Heisenberg uncertainty relations for relativistic bosons}
\author{Iwo Bialynicki-Birula}\email{birula@cft.edu.pl}
\affiliation{Center for Theoretical Physics, Polish Academy of Sciences\\
Aleja Lotnik\'ow 32/46, 02-668 Warsaw, Poland}
\author{Adam Prystupiuk}
\affiliation{Faculty of Physics, University of Warsaw, Pasteura 5, 02-093 Warsaw, Poland}
\begin{abstract}
This work completes the program started by I. Bialynicki-Birula and Z. Bialynicka-Birula [Uncertainty relation
for photons, \href{https://doi.org/10.1103/PhysRevLett.108.140401}{Phys. Rev. Lett. 108, 140401 (2012)}; Heisenberg uncertainty relation for photons, \href{https://doi.org/10.1103/PhysRevA.86.022118}{Phys. Rev. A
86, 022118 (2012)}; Heisenberg uncertainty relation for relativistic electrons, \href{https://doi.org/10.1088/1367-2630/ab3076}{New J. Phys. 21, 073036 (2019)}]
to derive the Heisenberg uncertainty relation for relativistic particles. Sharp uncertainty relations for massive
relativistic particles with spin 0 and spin 1 are derived. The main conclusion is that the uncertainty relations
for relativistic bosons are markedly different from those for relativistic fermions. The uncertainty relations for
bosons are based on the energy density. It is shown that the uncertainty relations based on the time component
of the four-current, as we have done previously for electrons, are untenable because they lead to contradictions. 
DOI: \href{https://doi.org/10.0.4.79/PhysRevA.103.052211}{10.1103/PhysRevA.103.052211}
\end{abstract}

\maketitle

\section{Introduction}

In this work, we complete our investigation of the Heisenberg-type uncertainty relations for relativistic particles. In previous papers, we derived the uncertainty relations for photons \cite{bb1,bb2} and also for relativistic spin-1/2 particles \cite{bb3}.

\begin{figure}
\centering
\includegraphics[width=0.35\textwidth,height=0.3\textheight]
{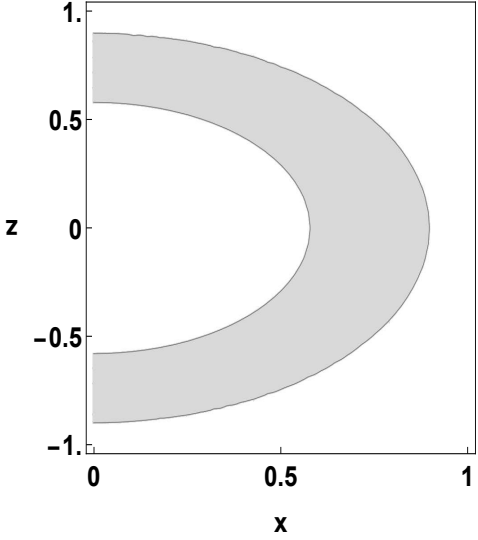}
\caption{The contour lines of charge density $\rho$  as a function of $x$ and $z$ plotted for $y=0$. In the shaded areas the charge density is negative. Owing to the rotational symmetry of the solution, in the three-dimensional picture we have spherical shells centered at the origin. In three dimensions inside each shell the charge density is negative. The parameters used to produce this plot are: $m=1,\,a=0.5$ and $t=0.05$.}\label{fig1}
\end{figure}

The purpose of this work is to show first that for relativistic massive bosons the uncertainty relation based on the charge density $\rho$, as has been done in \cite{bb3}, is unacceptable. This has been noticed already by Bjorken and Drell \cite{bd}. The argument will be presented for spin-0, in which case $\rho$ has the form
\begin{equation}\label{cur}
\rho=\frac{i}{2}\left(\phi^*\partial_t\phi
-\phi\partial_t\phi^*\right).
\end{equation}
For spin-1 the argumentation proceeds along similar lines. To prove our assertion, we consider the following solutions of the wave equation for massive spin 0 particles, the Klein-Gordon equation, $(\hbar=1,\,c=1)$:
\begin{align}\label{sol}
\phi&=\frac{1}{4\pi}\int\!d^3p f(p) e^{-(a+it)\sqrt{m^2+p^2}+i{\bm p}\cdot{\bm r}}\nonumber\\
&=\frac{1}{r}\int_0^\infty\!\!dp p\sin(pr)f(p)e^{-(a+it)\sqrt{m^2+p^2}}.
\end{align}
These integrals cannot be evaluated analytically but the numerical integration leads to the conclusion that the time component of the four-current cannot serve as a representation of the probability distribution because it is not positive definite. In Fig.~\ref{fig1} we show the results for $f(p)=\cos(p/m)/\sqrt{m^2+p^2}$. In the shaded region $\rho$ is negative. The dimensional quantities in the figure are measured in natural units based on $\hbar,\,c$ and $m$. Negative values disqualify $\rho$ as a measure of the probability distribution for a particle in space. 

This result forces us to replace, in the formulation of the uncertainty relation, the time component of the current with the energy density $\epsilon({\bm r})$.

It is worthwhile to mention at this point that the uncertainty relation based on the energy density \cite{wp} is, in turn, unacceptable for spin-1/2 particles, as was shown in \cite{bbc}. In Sec. \ref{section-2} we derive the uncertainty relation based on the energy density for the spin-0 particles, and in Sec. \ref{section-3} we do it for the spin-1 particles.

\section{Position-momentum uncertainty relation for spin 0 particles}\label{section-2}

The energy density for spin-0 particles is \cite{lsb,sw}
\begin{equation}\label{en0}
\epsilon({\bm r})=\pi^*({\bm r})\pi({\bm r})+\bm{\nabla}\phi^*({\bm r})\cdot\bm{\nabla}\phi({\bm r})+m^2\phi^*({\bm r})\phi({\bm r}),
\end{equation}
where $\pi({\bm r})=\dot{\phi}({\bm r})$. This expression is clearly positive definite so that, in contrast to $\rho$, it can be used to formulate the uncertainty relation. We consider only the wave function describing the particle (positive frequency part). The time dependence will not be shown explicitly since the uncertainty relation is always expressed at a fixed time. As we have done in our earlier publication, we will conduct the variational analysis in the momentum representation because it greatly simplifies the calculations. To this end, we represent $\phi$ and $\pi$ at $t=0$ in the form
\begin{subequations}
\begin{align}\label{r}
\phi({\bm r})&=\int\!\!\frac{d^3p}{\sqrt{2}(2\pi)^{3/2}E_p}e^{i{\bm p}\cdot{\bm r}}f({\bm p}),\\
\pi({\bm r})&=-i\int\!\!\frac{d^3p}{\sqrt{2}(2\pi)^{3/2}}e^{i{\bm p}\cdot{\bm r}}f({\bm p}),
\end{align}
\end{subequations}
where $E_p=\sqrt{m^2+p^2}$.
The norm and the dispersion of momentum expressed in terms of $f({\bm p})$ are
\begin{equation}\label{norm}
N^2=\int\!\!d^3p\,f^*({\bm p})f({\bm p})
\end{equation}
and
\begin{equation}\label{dp2}
\Delta p^2=
\frac{1}{N^2}\int d^3p\,{\bm p}^2f^*({\bm p})f({\bm p}).
\end{equation}

The dispersion $\Delta r^2$ based on the energy density is
\begin{equation}\label{dr}
\Delta r^2=\frac{1}{N^2}\int\!\!d^3r\,{\bm r}^2\epsilon({\bm r}).
\end{equation}
We assumed that the origin of the coordinate system is at the center of energy. In order to express $\Delta r^2$ in momentum representation, we use the formulas
\begin{subequations}
\begin{align}\label{r1}
{\bm r}\phi({\bm r})&=i\int\!\!\frac{d^3p}{\sqrt{2}(2\pi)^{3/2}}e^{i{\bm p}\cdot{\bm r}}{\bm\partial}\frac{f({\bm p})}{E_p},\\
{\bm r}\pi({\bm r})&=\int\!\!\frac{d^3p}{\sqrt{2}(2\pi)^{3/2}}e^{i{\bm p}\cdot{\bm r}}{\bm\partial}f({\bm p}),
\end{align}
\end{subequations}
where $\bm{\partial}$ denotes the gradient in momentum space. With the help of these formulas, after the integration over $\bm r$, the dispersion of the position takes on the form
\begin{widetext}
\begin{align}\label{dr0}
\Delta r^2&=\frac{1}{2N^2}
\int\!\!d^3p\left[{\bm\partial}(f^*({\bm p}))
\!\cdot\!{\bm\partial}(f({\bm p}))+\sum_k\partial_k \frac{\bm{p}f^*({\bm p})}{E_p}
\cdot\partial_k\frac{\bm{p}f({\bm p})}{E_p}+m^2{\bm\partial} \frac{f^*({\bm p})}{E_p}
\cdot{\bm\partial}\frac{f({\bm p})}{E_p}\right]\nonumber\\
&=\frac{1}{N^2}
\int\!\!d^3p\left[{\bm\partial}(f^*({\bm p}))
\cdot{\bm\partial}(f({\bm p}))+\left( \frac{m^2}{2E_p^4}+\frac{1}{E_p^2}\right)f^*({\bm p})f({\bm p})\right],
\end{align}
\end{widetext}

In order to derive the uncertainty relation, we must find the lowest value of $\gamma^2=\Delta r^2\Delta p^2$. This will be done with the use of the variational procedure.

\begin{figure}[t]
\begin{center}
\includegraphics[width=0.4\textwidth,height=0.2\textheight]
{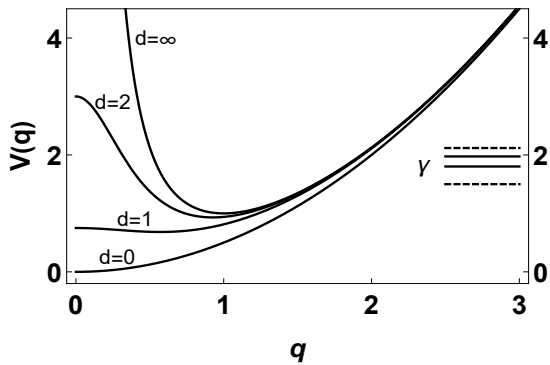}
\caption{The potential $V(q)$ plotted for four values of $d$. All quantities are measured in natural units based on $\hbar,\,c$, and $m$. The horizontal lines show the energy levels. The dashed lines represent the exact values and the values represented by solid lines were obtained by numerical integration of the eigenvalue equation. }\label{fig2}
\end{center}
\end{figure}

The variation of $\gamma^2$ with respect to $f^*({\bm p})$ can be calculated using the Leibniz rule for the variational calculus:
\begin{align}\label{leib}
\frac{\delta \gamma^2}{\delta f^*({\bm p})}&=\frac{\delta \Delta r^2}{\delta f^*({\bm p})} \Delta p^2 +\frac{\delta \Delta p^2}{\delta f^*({\bm p})} \Delta r^2=0.
\end{align}
Therefore
\begin{equation}\label{leib1}
\Big[\Delta p^2\left(\!\!-\Delta_p+\frac{m^2}{2E_p^4}+\frac{1}{E_p^2}\right)+
p^2\Delta r^2-2\gamma^2\Big]f(\bm{p})=0.
\end{equation}
This leads to the following eigenvalue equation:
\begin{equation}\label{eig}
\left[-\frac{1}{2}\Delta_q+\frac{d^2/2}{1+d^2q^2}
+\frac{d^2/4}{(1+d^2q^2)^2}+\frac{q^2}{2}\right]f(\bm q)=\gamma f(\bm q),
\end{equation}
where ${\bm q}$ is the rescaled momentum and $d$ is the dimensionless parameter introduced in \cite{bb3},
\begin{equation}\label{d}
{\bm q}=\frac{{\bm p}}{mcd},\quad
d=\frac{1}{mc}\left(\frac{\hbar^2\Delta p^2}{\Delta r^2}\right)^{1/4}.
\end{equation}
This equation may be viewed as the eigenvalue equation for a particle in the potential $V(q)$ of a modified harmonic oscillator.
\begin{align}\label{ee}
V(q)=\frac{d^2/2}{1+d^2q^2}+\frac{d^2/4}{(1+d^2q^2)^2}+q^2/2.
\end{align}
The dependence of the potential $V(q)$ on $d$ is illustrated in Fig.~\ref{fig2}.
From Eq. (\ref{eig}), we can see that the angular dependency of functions $f(\bm{q})$ contributes to the potential by adding the centrifugal term $l(l+1)/q^2$. The increase in potential will, in turn, increase the dispersions. Spin-0 particles have no distinguished direction; hence it makes sense that the solution exhibit spherical symmetry. Therefore we consider only those functions $f(\bm{q})$ that depend on the length of momentum vector.

This eigenvalue equation has no analytic solutions for an arbitrary value of $d$ but the eigenfunctions can be found in two limiting cases $m=\infty$ and $m=0$, i.e., for $d=0$ and $d=\infty$:
\begin{equation}\label{ef}
f_\infty(q)=e^{-q^2/2},\quad f_0(q)=q^{\sqrt{5}/2-1/2}e^{-q^2/2}.
\end{equation}
The first case gives the nonrelativistic result $\gamma=3/2$. The second case gives $\gamma=1+\sqrt{5}/2$. The same result was obtained in Ref. \cite{bb2} for photons and in Ref.  \cite{bb3} for massless spin-1/2 particles.
\section{Position-momentum uncertainty relation for spin 1 particles}\label{section-3}

The uncertainty relation for spin-1 particles is more complicated because the lowest value of $\gamma$ depends on the choice of the direction of the vector field.
The Lagrangian density of the vector field is given by the formula
\begin{equation}
\mathcal{L}(\bm{r})=-\frac{1}{4} f^*_{\mu\nu}(\bm{r})f^{\mu\nu}(\bm{r})
+\frac{1}{2}m^2A^*_{\mu}(\bm{r})A^{\mu}(\bm{r}),
\end{equation}
where $f_{\mu\nu}(\bm{r})=\partial_\mu A_\nu(\bm{r})-\partial_\nu A_\mu(\bm{r})$. We allowed the vector bosons to carry charge. The spatial components of field $A_\mu(\bm{r})$ will be denoted as $\bm{\phi}(\bm{r})$, while the canonically conjugate field is denoted as $\bm{\pi}^*(\bm{r})=\frac{\partial \mathcal{L}}{\partial \bm{\dot{\phi}}(\bm{r})}$.
The counterpart of Eq. (\ref{en0}) can now be expressed as \cite{qed,gr},
\begin{align}\label{en1}
\epsilon({\bm r})= &\bm{\pi}^*({\bm r})\cdot\bm{\pi}({\bm r})+m^{-2}\bm{\nabla}\!\cdot\!\bm{\pi}^*({\bm r})\bm{\nabla}\!\cdot\!\bm{\pi}({\bm r})\nonumber\\
&+\bm{\nabla}\times\bm{\phi}^*({\bm r})\!\cdot\!\bm{\nabla}\times\bm{\phi}({\bm r})+m^2\bm{\phi}^*({\bm r})\!\cdot\!\bm{\phi}({\bm r}),
\end{align}
where the vectors fields $\bm{\phi}({\bm r})$ and $\bm{\pi}({\bm r})$ satisfy the equations of motion generated by the Hamiltonian $\int\!\!d^3r\,\epsilon({\bm r})$,
\begin{subequations}
\begin{align}\label{reqm}
\frac{d}{dt}\bm{\phi}({\bm r})&=\bm{\pi}({\bm r})-m^{-2}\bm{\nabla}\left[\bm{\nabla}
\!\cdot\!\bm{\pi}({\bm r})\right],\\
\frac{d}{dt}\bm{\pi}({\bm r})&=-\bm{\nabla}\times\left[\bm{\nabla}
\times\bm{\phi}({\bm r})\right]-m^2\bm{\phi}({\bm r}).
\end{align}
\end{subequations}
These equations lead to the following connection between the Fourier representations of $\bm\phi$ and $\bm\pi$,
\begin{equation}\label{conn}
-iE_p\tilde{\bm\pi}({\bm p})={\bm p}\times\left({\bm p}
\times\tilde{\bm \phi}({\bm p})\right)
-m^2\tilde{\bm \phi}({\bm p}).
\end{equation}

In principle, we should search for the minimal value of $\gamma$ among all functions $\tilde{\bm \phi}({\bm p})=\{f_x({\bm p}),f_y({\bm p}),f_z({\bm p})\}/E_p$. In general, the variational method produces three complicated coupled equations for the functions $f_i({\bm p})$. We shall not write down these equations in the most general case, because to find the lowest value of $\gamma$ it is sufficient to consider two limiting cases:  $m=\infty$ and $m=0$ . The first case reproduces the standard nonrelativistic result, while the second case gives our previous result for photons \cite{bb2}.

To obtain the nonrelativistic limit, we choose the direction of the polarization vector along the $z$ axis:
\begin{align}
\bm{\phi}(\bm{r})=\int \frac{d^3p}{\sqrt{2}(2\pi)^{3/2}}
\frac{f(\bm{p})e^{i\bm{p}\cdot\bm{r}}}
{\sqrt{2m^2+p_x^2+p_y^2}}\left[
\begin{array}{c}0\\0\\1
\end{array}\right].
\end{align}
The denominator $\sqrt{2m^2+p_x^2+p_y^2}$  was introduced to make the formulas for $N^2$ and $\Delta p^2$ the same as for the spin-0 case given by  Eqs. (\ref{norm}) and (\ref{dp2}). The variation of $\gamma^2=\Delta p^2 \Delta r^2$ with respect to $f^*(\bm{p})$ leads to a fairly complicated equation. In order to obtain the lowest bound for $\gamma$, we will consider only the two limiting cases $m=\infty$ and $m=0$. The formulas for $\Delta r^2$ are
\begin{widetext}
\begin{align}
m=\infty\qquad \Delta r^2&=\frac{1}{N^2}
\int d^3p\left[{\bm\partial}(f^*({\bm p}))
\cdot{\bm\partial}(f({\bm p}))\right],\label{dr3a}\\
m=0\qquad \Delta r^2&=\frac{1}{N^2}
\int d^3p\left[{\bm\partial}(f^*({\bm p}))
\cdot{\bm\partial}(f({\bm p}))+\frac{1}{p_x^2+p_y^2}
f^*({\bm p})f({\bm p})\right].\label{dr3b}
\end{align}

The first formula coincides with that for spin 0 when in (\ref{dr0}) we put $m\to\infty$. Hence for spin 1 we also obtain the nonrelativistic limit $\gamma=3/2$. In the massless case we would expect a greater value due to the angular dependence in (\ref{dr3b}). Indeed, the variational equation can be solved, and it gives $\gamma=5/2$ and $f(q)=qe^{-5q^2/4}$ where $q^2=p^2/\Delta p^2$.

In order to obtain the massless limit, we choose the polarization vector in the direction of the momentum vector $\bm{p}/\abs{\bm p}$. This gives the following form of $\bm{\phi}$ and $\bm{\pi}$ at $t=0$:
\begin{equation}
\bm{\phi}(\bm{r})=\int \frac{d^3p}{\sqrt{2}(2\pi)^{3/2}}\frac{\bm{p}}{m\abs{\bm{p}}}f(\bm{p})e^{i\bm{p}\cdot\bm{r}},
\end{equation}
\begin{equation}
\bm{\pi}(\bm{r})=-i \int \frac{d^3p}{\sqrt{2}(2\pi)^{3/2}}\frac{m \bm{p}}{E_p \abs{\bm{p}}} f(\bm{p})e^{i\bm{p}\cdot\bm{r}}.
\end{equation}
The norm and the dispersion of momentum again have the same form as in Eqs. (\ref{norm}) and (\ref{dp2}) for spin 0, but the dispersion of position is markedly different:
\begin{align}\label{dr1}
\Delta r^2&=\frac{1}{2N^2}
\int d^3p\left[{\bm\partial} \frac{\abs{\bm{p}}f^*({\bm p})}{E_p}
\cdot{\bm\partial}\frac{\abs{\bm{p}}f({\bm p})}{E_p}+m^2\sum_k\partial_k \frac{\bm{p}f^*({\bm p})}{\abs{\bm{p}}E_p}
\cdot\partial_k\frac{\bm{p}f({\bm p})}{\abs{\bm{p}}E_p}+{\bm\partial}\frac{\bm{p}f^*({\bm p})}{\abs{\bm{p}}}
\cdot{\bm\partial}\frac{\bm{p}f({\bm p})}{\abs{\bm{p}}}\right]\nonumber\\
&=\frac{1}{N^2}
\int d^3p\left[{\bm\partial}(f^*({\bm p}))
\cdot{\bm\partial}(f({\bm p}))+\left(\frac{1}{\abs{\bm{p}}^2}
+\frac{m^2}{\abs{\bm{p}}^2E_p^2}+ \frac{m^2}{2E_p^4}\right)f^*({\bm p})f({\bm p})\right].
\end{align}
In the general case, the variation of $\gamma^2=\Delta p^2 \Delta r^2$ with respect to $f^*(\bm{p})$ leads to a fairly complicated equation,
\begin{equation}\label{var1}
\frac{1}{2}\left[-\frac{1}{q^2}\partial_q(q^2\partial_q)
-\frac{1}{q^2\sin\theta}
\partial_\theta(\sin\theta\partial_\theta)
-\frac{1}{q^2\sin^2\theta}\partial_\varphi^2
+V(q)\right]f(q,\theta,\varphi)=\gamma f(q,\theta,\varphi),
\end{equation}
\end{widetext}
where the potential now has the form
\begin{equation}\label{pot2}
V(q)=q^2+\frac{1}{q^2}+\frac{1}{q^2(1+q^2d^2)}
+\frac{d^2}{2(1+d^2q^2)^2}.
\end{equation}
Equation (\ref{var1}) allows for the separation of variables ($m$ stands here for the magnetic quantum number),
\begin{equation}
f(q,\theta,\varphi)=g(q)h(\theta)e^{im\varphi},\label{sep}
\end{equation}
and the equations for the radial and angular parts are
\begin{align}
&\left[-\frac{1}{q^2}\partial_q(q^2\partial_q)
+\frac{j(j+1)}{q^2}+V(q)\right]g(q)=2\gamma g(q),\label{rad}\\
&\left[-\frac{1}{\sin\theta}
\partial_\theta(\sin\theta\partial_\theta)
+\frac{m^2}{\sin^2\theta}\right]h(\theta)
=j(j+1)h(\theta).\label{ang}
\end{align}
Solutions of the differential equation for the angular part are the associated Legendre polynomials $P_j^m(\cos\theta)$. The nonsingular solutions exist for integer values of $j$ and $m$ with the condition $j\geq m\geq0$. Similar to the spin-0 case, the centrifugal term in (\ref{rad}) leads to an increase of the potential, and consequently, the dispersions. Therefore in the search for the lowest value of $\gamma$, the case of $j>0$ is, once again, of no interest. In contrast to the spin-0 case, however, the potential (\ref{pot2}) increases with the increasing mass. In two limiting cases, we obtain
\begin{equation}
V_{m=\infty}(q)=\frac{2}{q^2}+q^2,\;\;
V_{m=0}(q)=\frac{1}{q^2}+q^2.
\end{equation}
In both cases, the radial differential equation has analytic solutions
\begin{equation}
g_{m=\infty}(q)=qe^{-q^2/2},\;\;
g_{m=0}(q)=q^{\sqrt{5}/2-1/2}e^{-q^2/2},
\end{equation}
with the eigenvalues $\gamma_{m=\infty}=5/2$ and $\gamma_{m=0}=1+\sqrt{5}/2$. Thus for $m=\infty$ we obtain the result 5/2, which is larger than the nonrelativistic value of 3/2. In the massless case we obtain the same result as was obtained before for photons, massless spin-0, and massless spin 1/2 particles.
\vspace{0.1cm}
\section{Conclusions}
We completed the analysis of Heisenberg uncertainty relations for relativistic particles. The general conclusion is that relativistic corrections increase the lower bound for $\gamma^2=\Delta p^2\Delta r^2$. For all spins, the lowest value $\gamma=3/2\hbar$ is obtained in the nonrelativistic limit. The highest value $\gamma=(1+\sqrt{5}/2)\hbar$ is obtained for massless particles, i.e. in the ultrarelativistic case.

\end{document}